\documentclass{PoS}
\newcommand{\be}{\begin {eqnarray}}
\newcommand{\ee}{\end{eqnarray}}

\newcommand{\p}{\pi_{\Gamma}}
\newcommand{\s}{\sigma}

\newcommand\redout{\bgroup\markoverwith{\textcolor{blue}{\rule[0.5ex]{6pt}{1.0pt}}}\ULon}
\title{Borici-Creutz fermions on 2-dim lattice}

\ShortTitle{Borici-Creutz fermions on 2-dim lattice}

\author{\speaker{Jishnu Goswami}%
\\
       Physics Department, Indian Institute of Technology Kanpur, Kanpur-208016, India\\
       E-mail: \email{jgoswami@iitk.ac.in}}

\author{Dipankar Chakrabarti\\
        Physics Department, Indian Institute of Technology Kanpur, Kanpur-208016, India\\
        E-mail: \email{dipankar@iitk.ac.in}}
\author{Subhasish Basak\\
        School of Physical Sciences, NISER, Jatni-752050, India\\
        E-mail: \email{sbasak@niser.ac.in}}

\abstract{
Minimally doubled fermion proposed by Creutz and Borici is a promising chiral fermion formulation on  lattice. In this work, we present  excited state mass spectroscopy for the meson bound states in Gross-Neveu model using  Borici-Creutz fermion. We also evaluate the effective   fermion mass as a function of coupling constant  which shows a chiral phase transition at strong coupling. The lowest lying meson in 2-dimensional QED is also  obtained using  Borici-Creutz  fermion. 
 }

\FullConference{34th International Symposium on Lattice Field Theory\\
         24-30 July 2016\\
         University of Southampton, UK}

\begin{document}
\section{Introduction}
%%%%%%%%%%%%%%%%%%%%%%%%%%%%%%%%%%%%%%%%%%%%%%%
%Chiral fermion formulation is always a challenging task on lattice and
 Minimally doubled fermions in recent times have drawn attention as promising lattice formulations of chiral fermion. 
%By no-go theorem, the number of fermions with chiral symmetry on a lattice  at least be  two. 
There are two minimally doubled fermion formulations, one is by
Karsten\cite{K} and Wilczek\cite{W}  and the other one was developed by Creutz\cite{creutz} and Borici\cite{borici}.  Both the formulations break the hypercubic symmetry on the lattice \cite{bedaque} and thus allow non-covariant counter terms. So, the important  question is how bad the effects of the symmetry breaking are in a numerical simulation. It was  shown that a consistent renormalizable theory for minimally doubled fermion can be constructed by fixing only three counter terms allowed by the symmetry and the counter terms for BC action at  one loop in perturbation theory have been evaluated \cite{capitani1,capitani2}.
But, till date, sufficient numerical studies of the minimally doubled fermions have not been done. The purpose of this work is to investigate  Borici-Creutz(BC) formulation numerically  in some  simple models. 
%The BC fermion formulation was motivated by the fact that electrons on  graphene lattice are described by a massless quasi-relativistic Dirac equation.
% The  BC fermion describes two chiral modes or the two flavors of chiral fermion  located at $(0,0,0,0)$ and at $(\frac{\pi}{2},\frac{\pi}{2},\frac{\pi}{2},\frac{\pi}{2})$.
It was shown that in presence of gauge background with integer-valued topological charge, BC action satisfies the Atiyah-Singer index theorem\cite{dc}. In \cite{GCB}, using BC fermion  we have shown a chiral phase transition in the Gross-Neveu model.  In this work, we evaluate the  meson spectroscopy of the Gross-Neveu model  as well as in 2D QED (QED$_2$) with BC fermion. 
Chiral and parity-broken(Aoki) phase structures of the Gross-Neveu model have been studied for Wilson  and Karsten-Wilczek fermions \cite{CKM,misumi}. 
%A lattice simulation of the Gross-Neveu model   using  the Wilson fermion was done by Korzec et al\cite{korzec}, where the recovery of chirally invariant Gross-Neveu model from a lattice model was studied.
%The semimetal-insulator phase transition on a graphene lattice with Thirring type four fermion interactions has been studied by Hands and collaborators\cite{hands} and the strong coupling analysis of  the tight-binding graphene model with Kekule distortion term has been done by Araki\cite{araki}. 
%The mass spectra  in different models in 2d are evaluated using the Borici-Creutz  minimally doubled fermion formulation. 
By  hybrid Monte Carlo(HMC) simulation  we investigate the excited state  spectrum of the lattice Gross-Neveu model.  
%From the slope of the correlators,   some preliminary estimates are obtained   and then  variational method is used to extract the meson masses. 
Study of the excited state spectroscopy in the Gross-Neveu model with Wilson fermion \cite{Danzer} found the
% In their work, authors of \cite{Danzer}, 
 ground state as the only bound state and the other excited states as scattering states. With the BC fermion in this work, we have obtained three states,  two of them are bound states (ground state and one excited state) and the third one appears to be a scattering state. %We also evaluate the fermion mass in the model which shows that it is consistent with a chiral phase transition at large coupling   observed in \cite{GCB}.
Next we investigate the meson mass spectrum in QED in two dimension.  QED$_2$ having confinement serves as a toy model for QCD and hence QED$_2$ or Schwinger model has been  studied to great extent on lattice (see \cite{Gutsfeld,Gattringer} and references therein).  Schwinger model using Hamiltonian formalism on lattice has been investigated in \cite{cichy}.  QED$_2$ also serves as a good toy model for numerical study of chiral  fermions. A 2-flavor
Schwinger model with light fermions have been studied with    dynamical  overlap fermion \cite{ Giusti, Hip}. 
%Here, we study the  model with the minimally doubled fermion, namely, the BC fermion.

\section{Spectroscopy of the Gross-Neveu model}
%%%%%%%%%%%%%%%%%%%%%%%%%%%%
%Though there are strong coupling analyses in the literature,  until now, the Gross-Neveu model with BC fermion has not been studied  numerically  on a lattice.   
The free BC action in 2D  is written as,
  \be
  S&=&\sum_{n}\left[\frac{1}{2}\sum_{\mu}\overline{\psi}_n\gamma_{\mu}(\psi_{n+\mu}-\psi_{n-\mu})-\frac{i r}{2}\sum_{\mu}\overline{\psi}_n(\Gamma-\gamma_{\mu})(2\psi_{n}-\psi_{n+\mu}-\psi_{n-\mu}) \right.\nonumber \\
 && \left. -i(2-c_{3})\overline{\psi}_{n}\Gamma\psi_{n}+m_0\overline{\psi}_{n}\psi_{n}\right]\label{CB},
  \ee
  where, $\mu=1,2$ and  $\Gamma=\frac{1}{2}(\gamma_{1}+\gamma_{2})$ satisfies  $\{\Gamma,\gamma_{\mu}\} =1$.
Including four-fermion interactions, the  Gross-Neveu model on  lattice is given by
\begin{eqnarray}
       S&=&\sum_{n}\left[\frac{1}{2}\sum_{\mu}\overline{\psi}_n\gamma_{\mu}(\psi_{n+\mu}-\psi_{n-\mu})-\frac{i r}{2}\sum_{\mu}\overline{\psi}_n(\Gamma-\gamma_{\mu})(2\psi_{n}-\psi_{n+\mu}-\psi_{n-\mu}) \right.\nonumber \\
        && \left. -i(2-c_{3})\overline{\psi}_{n}\Gamma\psi_{n}+m_0\overline{\psi}_{n}\psi_{n}-\frac{g^{2}}{2N}[(\overline{\psi}_{n}\psi_{n})^{2}+(\overline\psi_{n}i\Gamma\psi_{n})^{2}\right],\label{gnmodel}
        \ee
where,  $g$ is the coupling constant which we consider  the same for both  four point (scalar and vector) interactions and we set $r=1$ in our calculations. Since the parity is broken by the BC action, a counter term $c_3$ is added to it. Detailed discussion about the  $c_3$-term can be found in \cite{GCB}.
The action is rewritten explicitly in terms of the auxiliary fields as
 \begin{equation}
     S=\sum_{m,n}\overline{\psi}_{m}M_{mn}\psi_{n}+{\frac{N}{2g^{2}}}(\sigma^{2}+\pi_{\Gamma}^{2}),
   \end{equation}
   where $N$ is the number of flavors. The auxiliary fields  are
   \begin{eqnarray}
  \s=-\frac{g^2}{N}(\overline{\psi}\psi), ~~~~
  \p=-\frac{g^2}{N}(\overline{\psi}i\Gamma\psi).
\end{eqnarray}
%\subsection{Correlators }
%%%%%%%%%%%%
For meson mass spectrum calculation, we need to evaluate the correlators
\be
C_{ij}(t)=\langle O_i(t)O_j^\dagger(0)\rangle.
\ee
In absence of the orbital angular momentum in 2D, the interpolators ($O_i$) are labelled by  parity  and charge conjugation only. We need to choose  appropriate operators which have good overlaps with the low lying states.  For the meson spectroscopy,  we consider only the odd parity interpolators. %The even parity interpolators are not considered as they  do not have  exponential decay and correspond to condensates\cite{Danzer}. 
The even parity interpolators correspond to condensate\cite{Danzer} and are not considered.
 Under parity $\psi(x,t)\to \gamma_2 \psi(-x,t)$ and the odd parity  interpolators can be constructed with $\gamma_1$ or $\gamma_5$. Along with the  local source, one can also construct the interpolators with the fields at  different lattice sites shifted along the spatial direction ie., with $\psi(x\pm n,t)$ . If one considers a relative negative sign in between $\psi(x+n,t)$ and $\psi(x-n,t)$ then this corresponds to a derivative source which are found to be important for excited state spectroscopy\cite{Danzer,Dsource}.  %Combining the field operators at different lattice sites, many interpolators can be constructed but it was found in our numerical analysis that  they mostly couple to the ground state. % In\cite{Danzer},  a set of nine different interpolators were listed.
 Here we  list some of the parity odd interpolators for the GN model which we expect to couple to ground state as well as excited states:
\be
O_1(t)&=&{\overline \psi}(x,t)\gamma_5\psi(x,t) \nonumber\\
%O_2(t)&=&{\bar \psi}(x+n,t)\gamma_5\psi(x-n,t)\nonumber\\
O_2(t)&=&\frac{1}{4}\big(({\overline \psi}(x+m,t)-\overline{\psi}(x-m,t)\big)\gamma_5\big(\psi(x+n,t)-\psi(x-n,t)\big), (m=3, n=3) \nonumber\\
%O_3(t)&=&{\bar \psi}(x,t)\gamma_1\psi(x,t)\nonumber\\
%O_4(t)&=&{\bar \psi}(x,t)\gamma_5\big(\psi(x+n,t)+\psi(x-n,t)\big)\nonumber\\
%O_5(t)&=&{\bar \psi}(x,t)\gamma_1\big(\psi(x+n,t)+\psi(x-n,t)\big)\nonumber\\
O_3(t)&=&\frac{1}{4}\big(({\overline \psi}(x+m,t)-\overline{\psi}(x-m,t)\big)\gamma_5\big(\psi(x+n,t)-\psi(x-n,t)\big), (m=5, n=3)\label{interpol}\\
O_4(t)&=&\frac{1}{4}\big(({\overline \psi}(x+m,t)-\overline{\psi}(x-m,t)\big)\gamma_1\big(\psi(x+n,t)-\psi(x-n,t)\big), (m=4, n=3)\nonumber\\
O_5(t)&=&\frac{1}{4}\big(({\overline \psi}(x+m,t)+\overline{\psi}(x-m,t)\big)\gamma_1\big(\psi(x+n,t)-\psi(x-n,t)\big), (m=5, n=3),\nonumber
\ee
where sum over $x$ is implied in order  to have  zero momentum  states and $\gamma_5=i\gamma_1\gamma_2$. All the interpolators are odd under $C$-parity ($C=-1$). With different values of $m$ and $n$, we can have different interpolators,  they
%but the one that are found to couple with ground state as well as the excited states are for the values listed above in Eq.(\ref{interpol}), other interpolators 
do not couple to new states but only reproduce the similar results.
\begin{figure}[htbp]
\centering
\begin{minipage}[c]{0.98\textwidth}
\small{(a)}\includegraphics[width=6cm,clip]{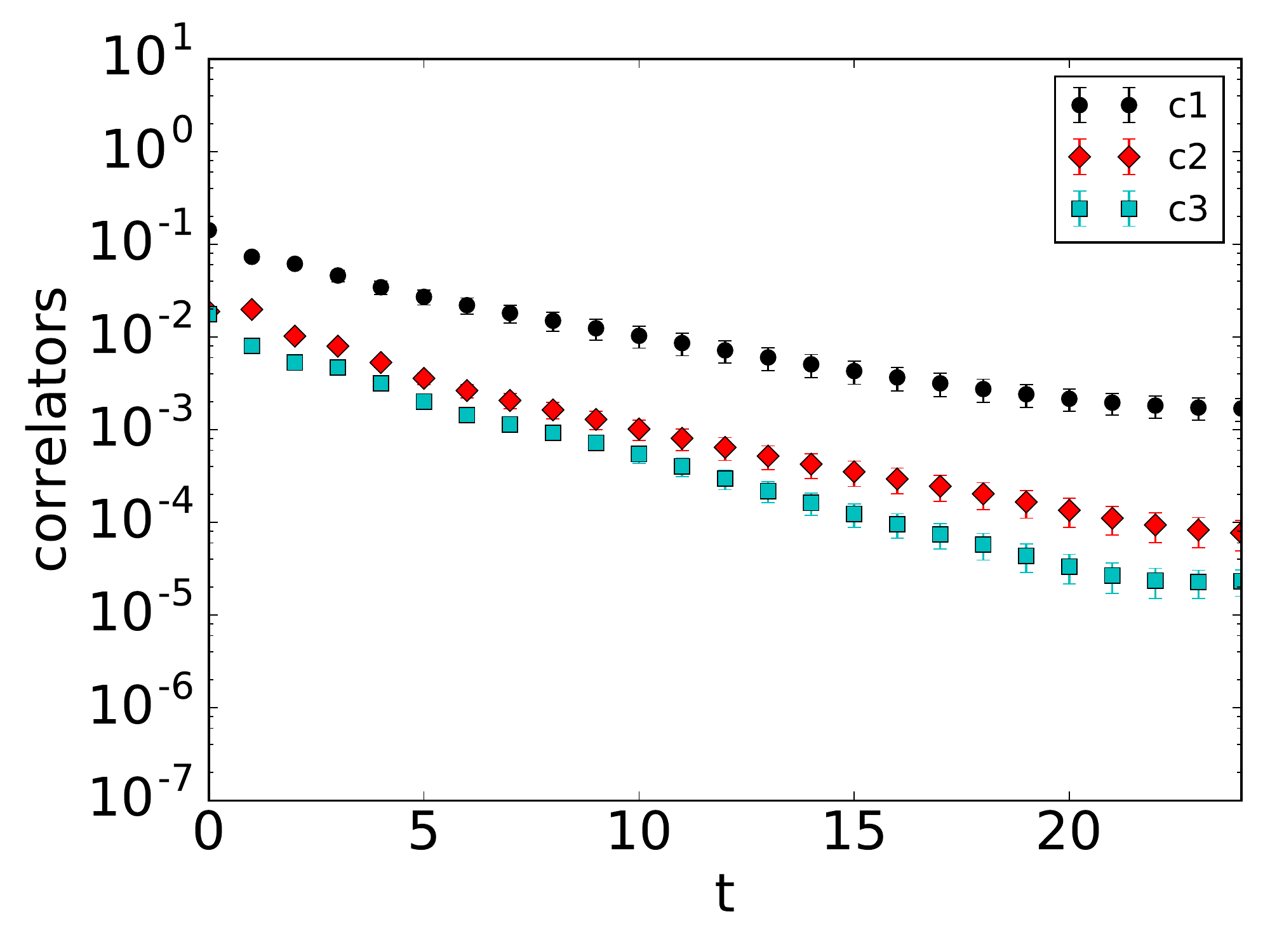}
\hspace{0.1cm}%
\small{(b)}\includegraphics[width=6cm,clip]{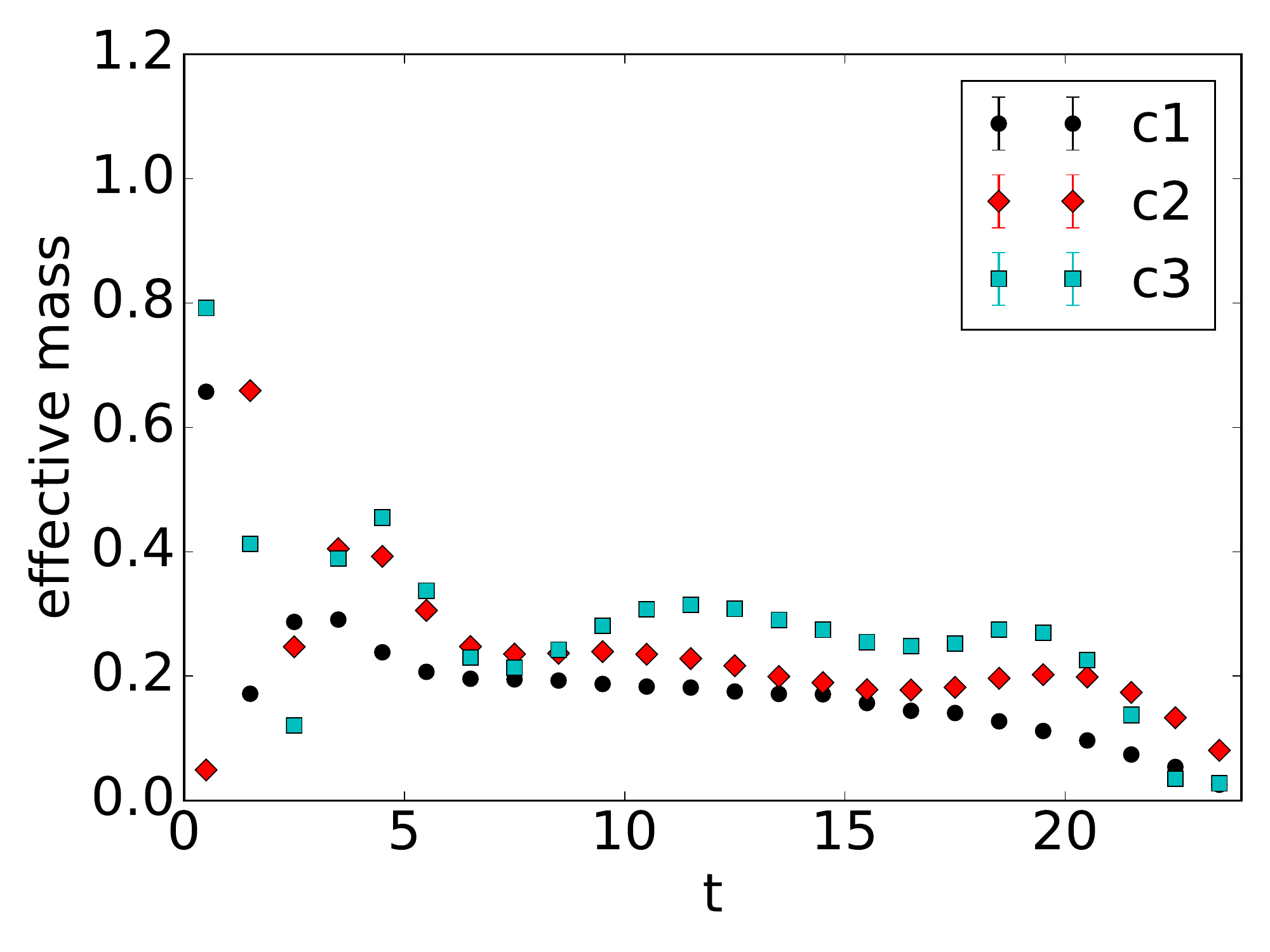}
\end{minipage}
\caption{ Diagonal correlators (in the plots $c1\equiv C_{11}$, etc.) %(in the lagend $cn=C_{nn}$ with interpolator $O_n$) 
and effective mass of meson in GN model for $16\times 48$ lattice\label{GN_meson_corr}}
\end{figure}
%\subsection{Effective mass calculation}
%%%%%%%%%%%%%%%%%%%%%
The effective masses are extracted from the correlators at different time slices by  the formula 
\be M_{eff}=\ln \left(\frac{c(t)}{c(t+1)}\right).\label{m_eff}
\ee
 The diagonal correlators $C_{ii}$ and corresponding effective masses are shown in Fig.\ref{GN_meson_corr}(a) and (b) for $m_0=0.03$ and $\beta=0.7$.  The small value of mass is taken to be close to the massless  limit.
%Eq.(\ref{m_eff})  is an approximate formula and found good for ground state but  can also produce approximate values for the excited states. 
As shown in Fig. \ref{GN_meson_corr}(b), except for the ground state, this procedure is not suitable to extract the excited states.   The variational method \cite{Michael,Luscher} provides a better picture for the excited state meson spectroscopy.
\begin{figure}[hptb]
\centering
\begin{minipage}[c]{0.98\textwidth}
\small{(a)}\includegraphics[width=6cm,clip]{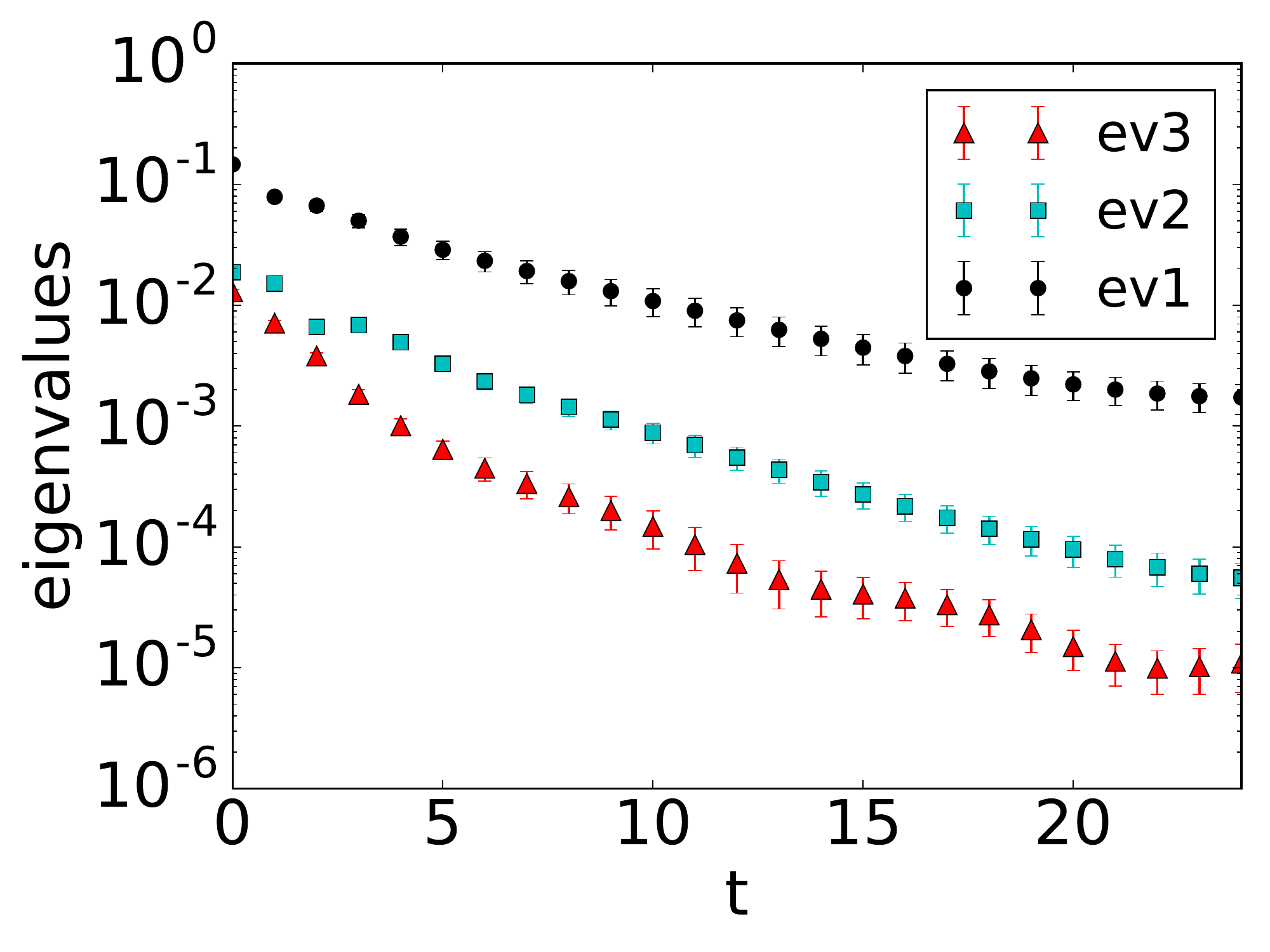}
\hspace{0.1cm}%
\small{(b)}\includegraphics[width=6cm, clip]{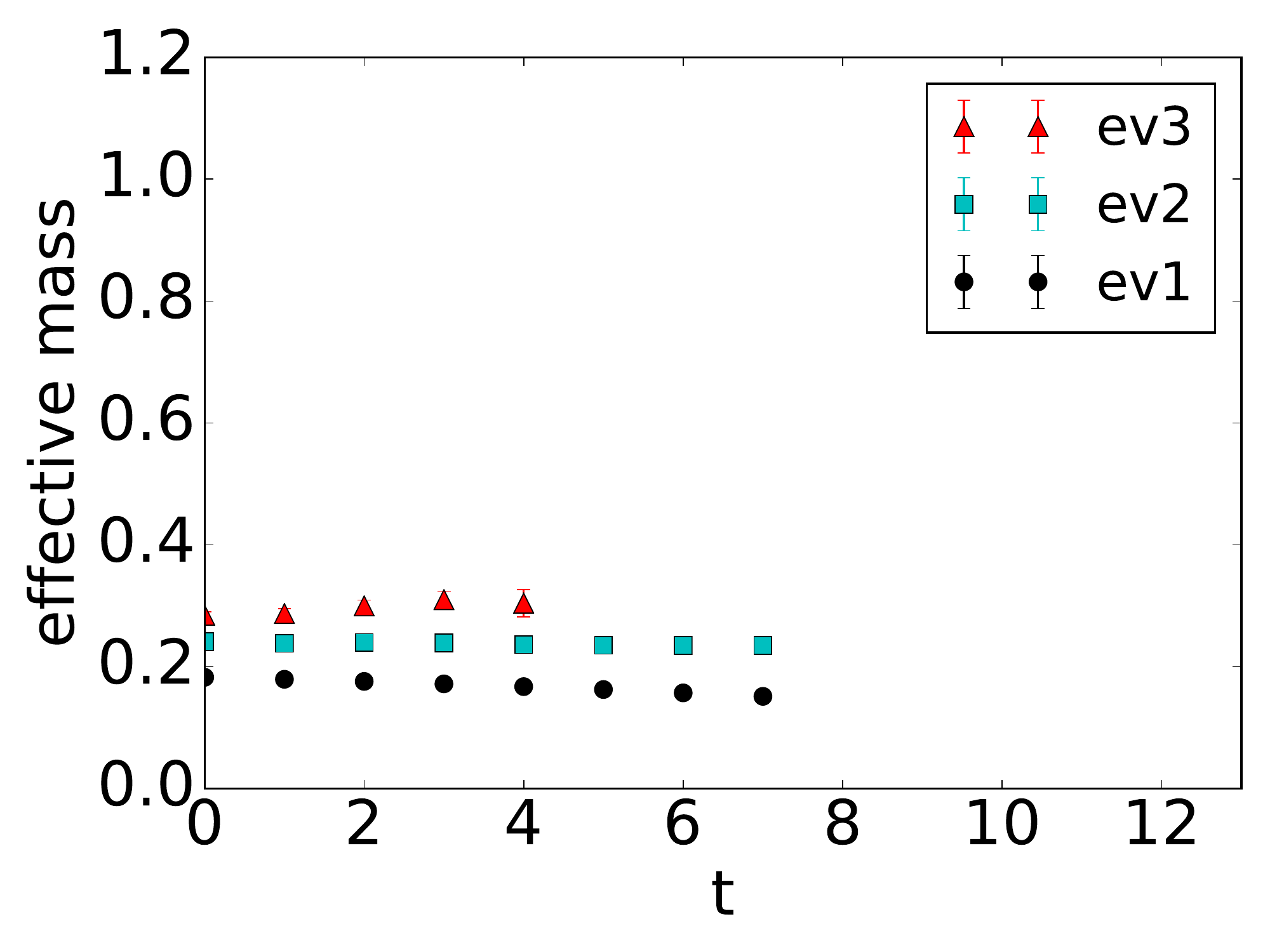}
\end{minipage}
%\caption{ Eigenvalues and effective mass of the correlators for $16 \times 48$ lattice  \label{GN_eigen_meff}}
%\end{figure}
%
%\begin{figure}[htbp]
\centering
\begin{minipage}[c]{0.98\textwidth}
\small{(c)}\includegraphics[width=6cm,clip]{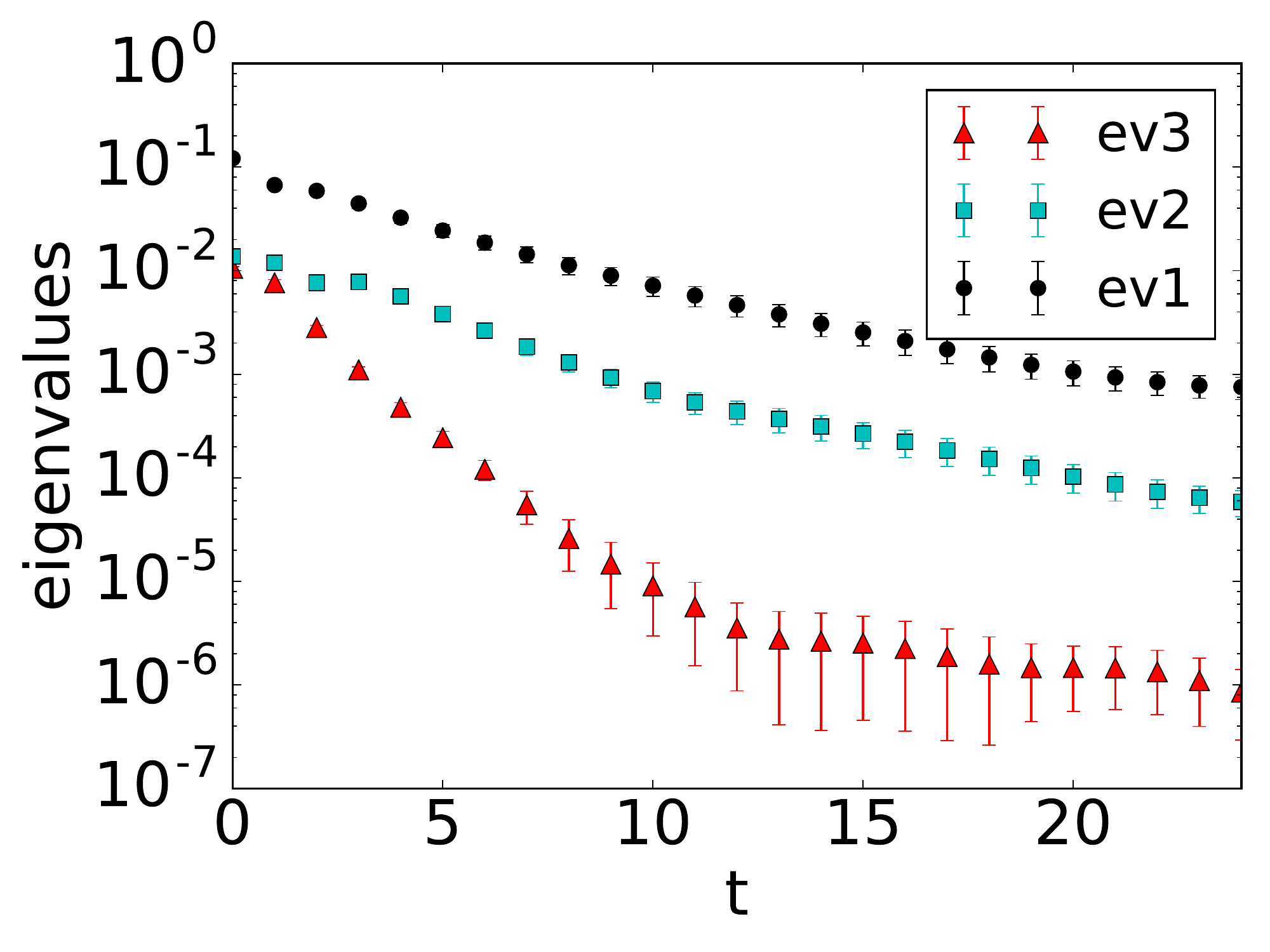}
\hspace{0.1cm}%
\small{(d)}\includegraphics[width=6cm,clip]{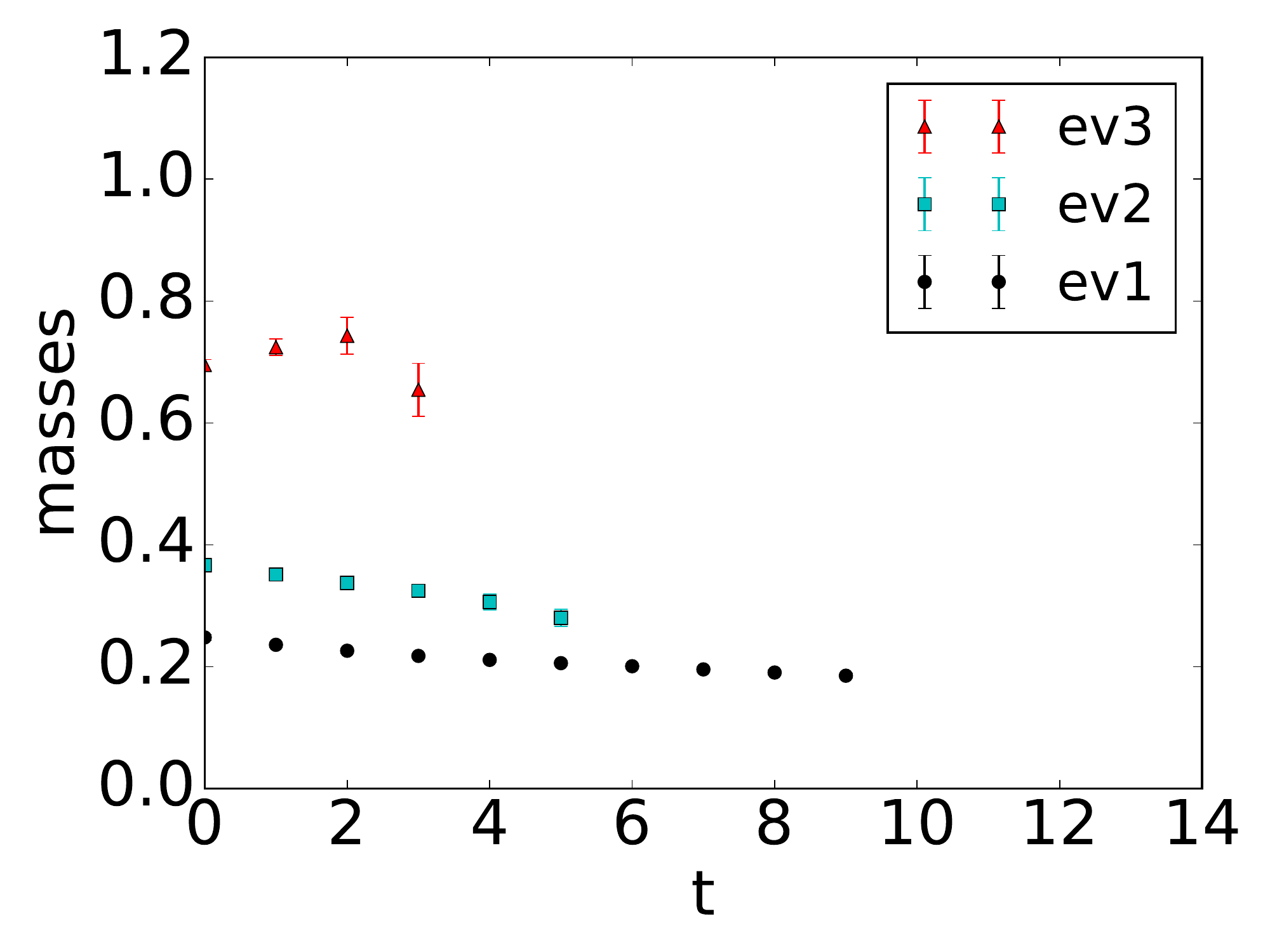}
\end{minipage}
%\caption{ Eigenvalues and effective mass of the correlators for $20 \times 48$ lattice  \label{GN_eigen_meff1}}
%\end{figure}
%
%\begin{figure}[htbp]
%\centering
%\begin{minipage}[c]{0.98\textwidth}
%\small{(e)}\includegraphics[width=5.6cm,clip]{eigenvalues_new_22.pdf}
%%\hspace{0.1cm}%
%\small{(f)}\includegraphics[width=5.6cm, clip]{mass_new_fit_vol22.pdf}
%\end{minipage}
%\caption{ Eigenvalues and effective mass of the correlators for $22 \times 48$ lattice  \label{GN_eigen_meff2}}
%\end{figure}
%
%\begin{figure}[htbp]
\centering
\begin{minipage}[c]{0.98\textwidth}
\small{(e)}\includegraphics[width=6cm,clip]{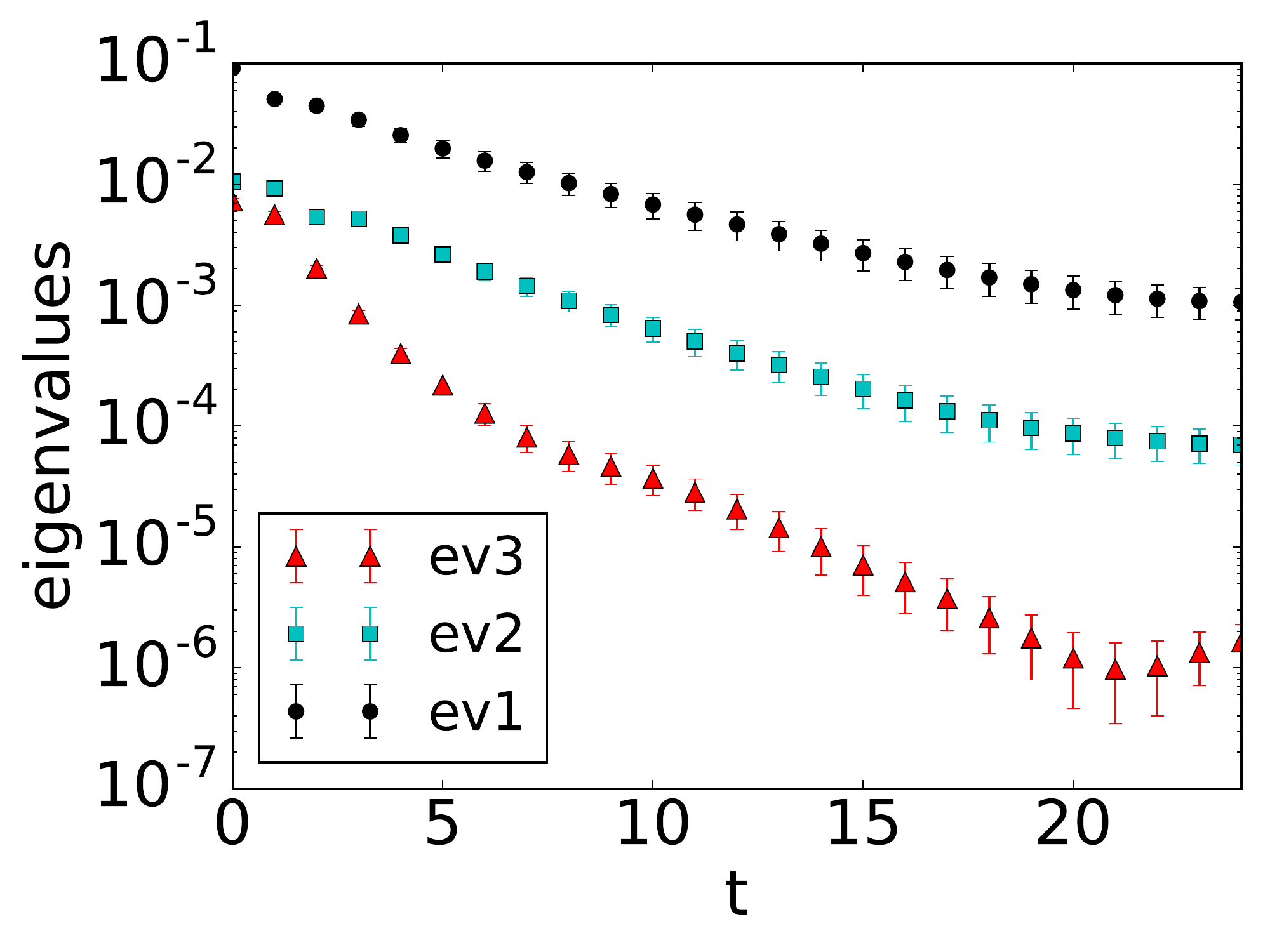}
\hspace{0.1cm}%
\small{(f)}\includegraphics[width=6cm, clip]{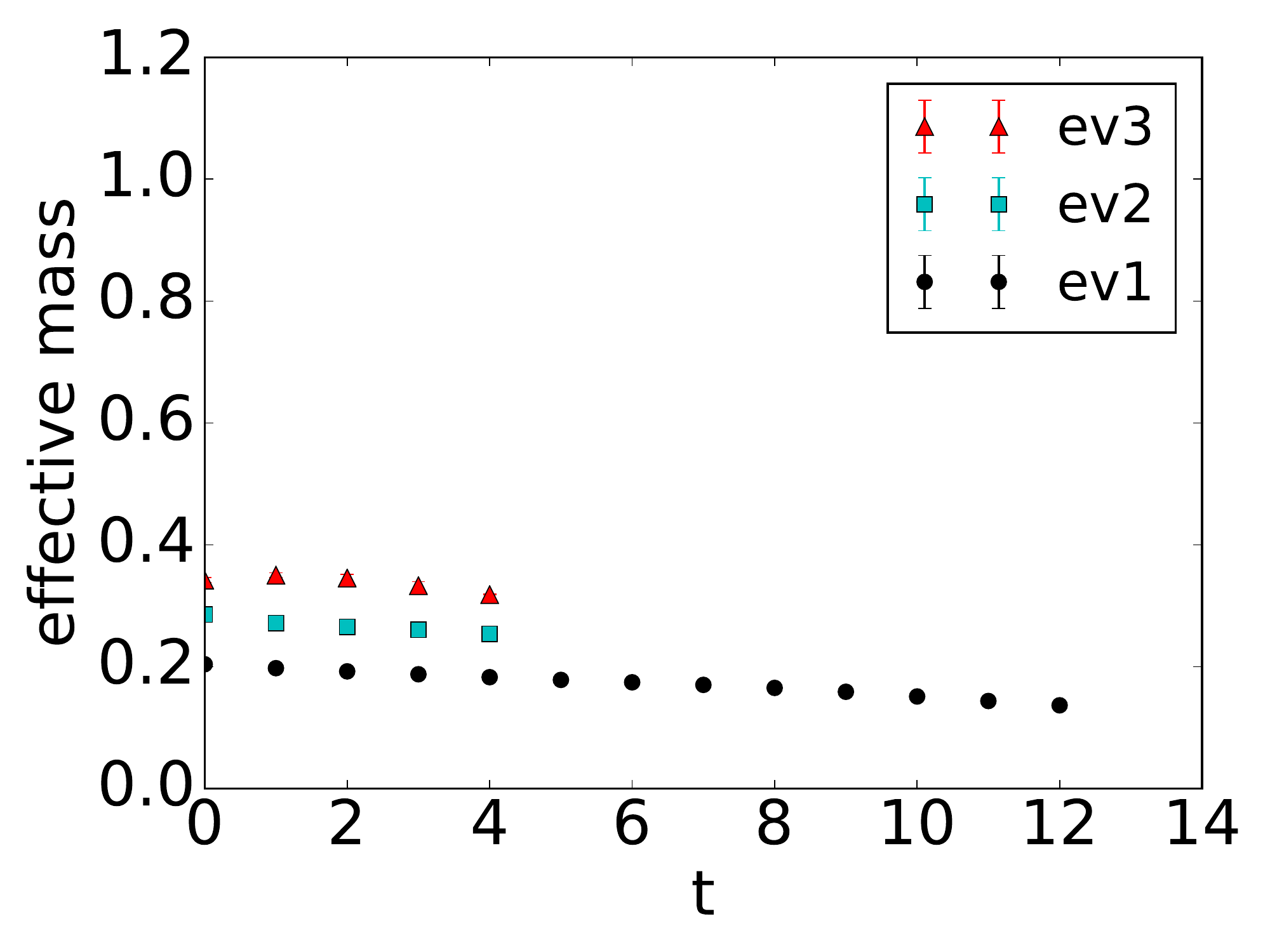}
\end{minipage}
\caption{ Eigenvalues and effective mass of the correlators.  (a) and (b) for  $16 \times 48$; (c) and (d) for $18 \times 48$; %(e) and (f)  for $22 \times 48$;  
and (e) and (f) for $24 \times 48$ lattices.  \label{GN_eigen_meff_all}}
\end{figure}
%%%%%%%%%%%%
\begin{figure}[htbp]
\centering
%\begin{minipage}[c]{0.98\textw
\includegraphics[width=6.5cm,clip]{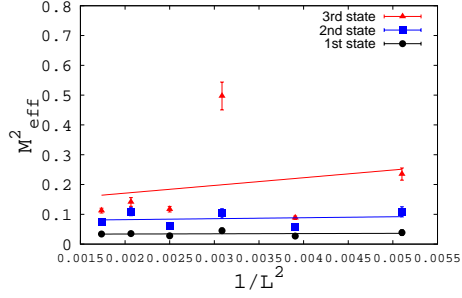}
\caption{\label{volume_dep} Volume dependence of the effective mass.  }
\end{figure}
  %to get the mass spectrum from eigenvalues
 In this method,  one solves the generalized eigenvalue problem %defined by
  \be
  C(t) \vec{v}^{(n)}=\lambda^{(n)}(t)C(t_0)\vec{v}^{(n)},
  \ee
  where $C(t)$ is the $N\times N$ correlation matrix  constructed from $N$ interpolators $O_i,~(i=1,2\cdots, N)$. The $n$-th eigenvalue behaves as
  \be
  \lambda^{(n)}(t)=e^{-(t-t_0)E_n}\Big[1+\mathcal{O}(e^{-(t-t_0)\Delta_n})\Big],
  \ee
  where $E_n$ is the energy of the $n$-th state and $\Delta_n$ is the energy gap between the neighboring states. 
In Fig.\ref{GN_eigen_meff_all} we show the eigenvalue and effective mass plots by solving the generalized eigen value problem for $16\times 48$,  $18 \times 48$, %~ 22\times 48$ 
and $24\times 48$ lattices with $O_1,~O_2$ and $O_3$ interpolators.  
We are unable to get any extra stable mass values by increasing the matrix dimension of the correlator basis and the results become noisier with more correlators. So, we present the results with only three correlators. %$O_1,~O_2$ and $O_3$. 
%The results become noisier if the matrix dimension is increased by including more correlators. 
%The eigenvalues and effective mass plots for
%the $16\times 48$ lattice  are shown in Fig.\ref{GN_eigen_meff}. Three mass plateau can be observed in Fig.\ref{GN_eigen_meff}(b). 
%In Fig.\ref{GN_eigen_meff_all} %, Fig.\ref{GN_eigen_meff2}  and Fig.\ref{GN_eigen_meff3} we have shown the eigenvalues and effective masses for $16\times 48$, $20 \times 48,~ 22\times 48$ and $24\times 48$ respectively. Three mass plateau are obtained for each lattice.
 In Fig.\ref{volume_dep}, the volume dependence of the effective masses are shown. The ground state and the first excited state show no  volume dependence and hence can  be considered as bound states. The second  excited state however shows  volume dependence. Specially, for $18\times 48$ lattice size, we get an anomalously large mass for the second excited state. The fit for the second excited state shown in Fig.\ref{volume_dep}  includes this anomalous point.  In general, scattering states show strong volume dependence and increase linearly with $1/L^2$, the volume dependence of the second excited state in our case is not very conclusive. %cannot be stated as neither a scattering state nor a bound state for sure. 
 But looking at the fit of the points we expect  it to be  a   scattering state. The results can be contrasted with \cite{Danzer}, where except the ground state, all the excited states show strong volume dependence and  are  scattering states.
\section{Meson in 2D QED}
%%%%%%%%%%%%%%%%%%%%%
In this section, we extend the study of spectroscopy with BC fermion formulation to gauge theory.  For this purpose, we  implement the BC fermion in  a 2D $U(1)$ gauge theory  and  extract the meson masses.  QED in 2D is also a confined theory and serves as a good toy model for QCD. 
%The Gross-Neveu model, having a discrete chiral symmetry undergoes spontaneous breaking, but  in the massless Schwinger model, the chiral symmetry is continuous and cannot be spontaneously broken.
The lattice action with BC fermion reads, 
\be
S=\beta\sum_p[1-\frac{1}{2}(U_p+U^{\dagger}_p)]+\phi^{\dagger}(D^{\dagger}D)^{-1}\phi.
\ee  where $U_p$ is the Wilson Plaquette action with
 \be 
 U_p=U_{i,\mu}U_{i+\mu,\nu}U^{\dagger}_{i+\nu,\nu}U^{\dagger}_{i,\nu}.
 \ee where, $i$ is the site index and $\mu,\nu$ are the directions
and $D$ is the BC Dirac operator %defined in eqn.(\ref{bcdirac}).
 defined as
 \be 
  D_{mn}&=&\frac{1}{2}(\gamma_\mu+i(\Gamma-\gamma_\mu))U_{\mu}(n-\mu)\delta_{n,m+\mu}-\nonumber\\
  &&\frac{1}{2}(\gamma_\mu-i(\Gamma-\gamma_\mu))U^{\dagger}_{\mu}(n)\delta_{n,m-\mu}-((2-c_3)i\Gamma-m_0)\delta_{m,n}.
 \ee
 \begin{figure}[htbp]
 \centering
\begin{minipage}[c]{0.98\textwidth}
\small{(a)}\includegraphics[width=5.6cm,clip]{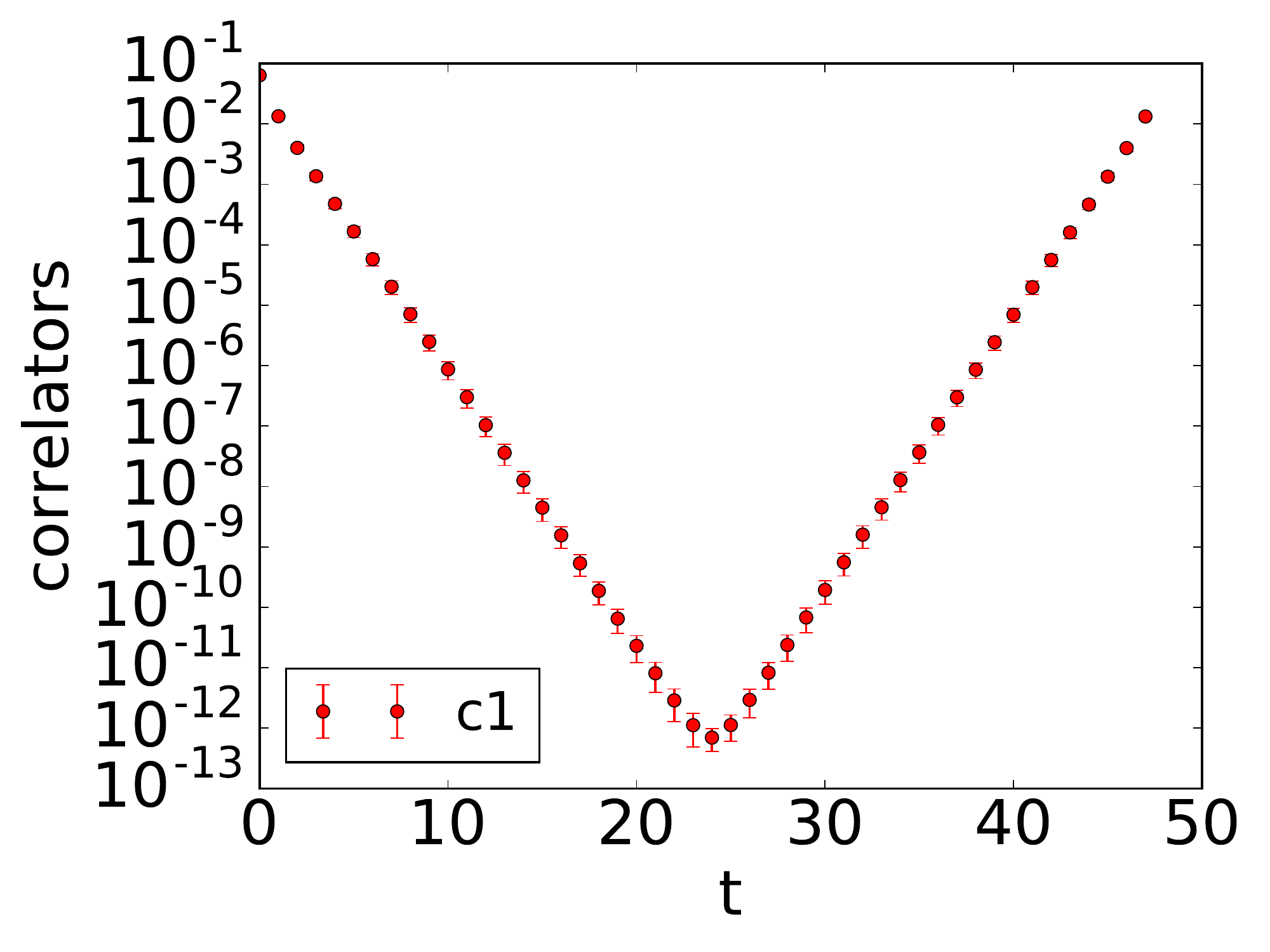}
\hspace{0.1cm}%
\small{(b)}\includegraphics[width=5.6cm,clip]{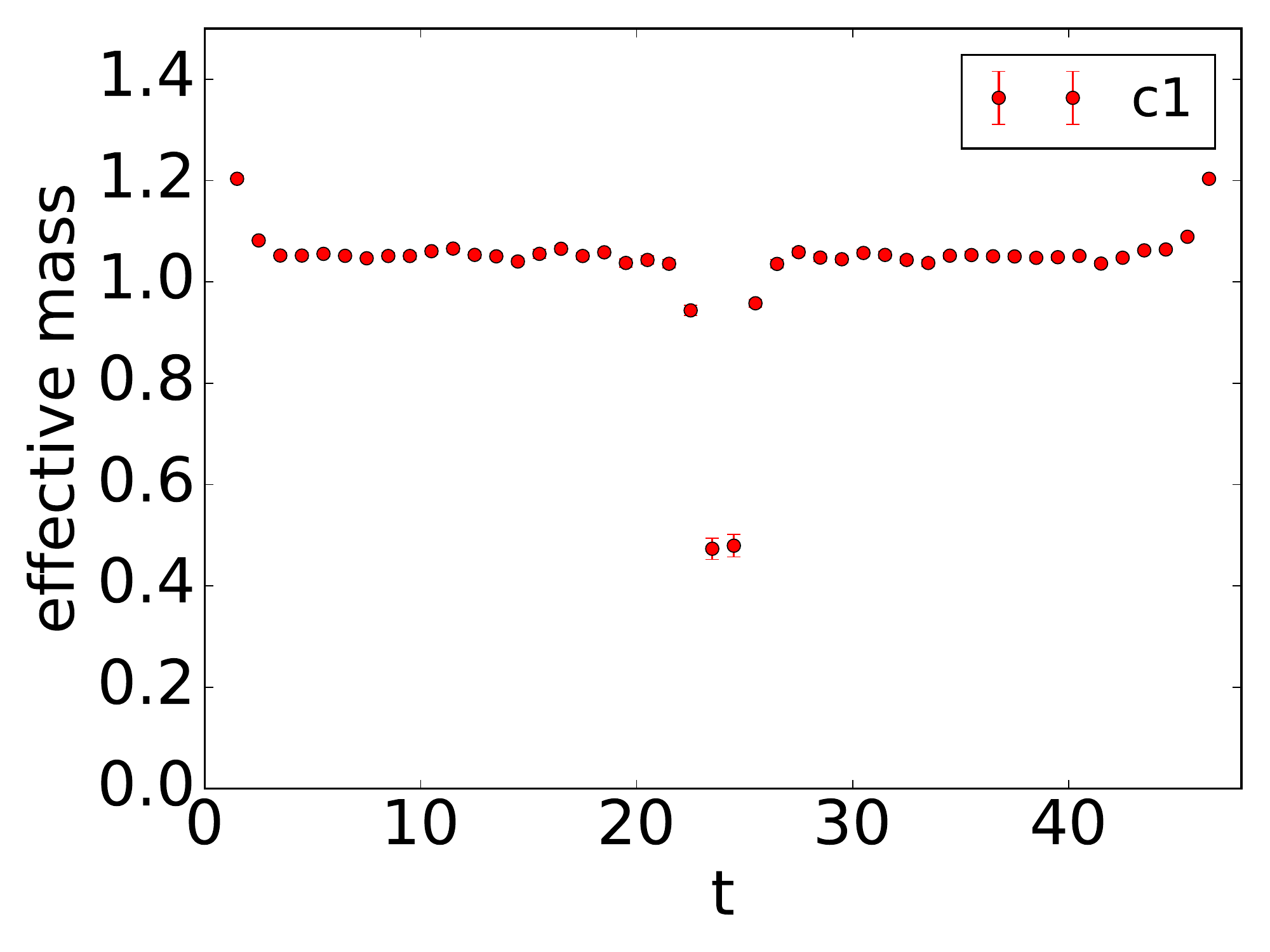}
\end{minipage}
\caption{Effective mass of meson in 2D QED for $m_0=0.05$ and $\beta=0.3$.\label{QED_meson}}
\end{figure}
\begin{figure}[htbp]
%\begin{minipage}[c]{0.98\textwidth}
%\small{(a)}
\centering
\includegraphics[width=6cm,clip]{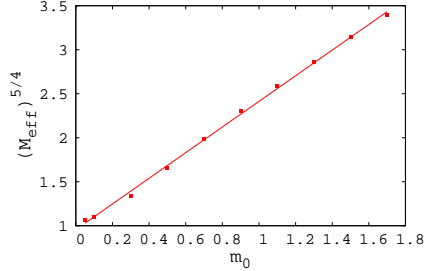}
%\hspace{0.1cm}%
%\small{(b)}\includegraphics[width=7cm, height=7cm,clip]{fig_QED_meson_mass.pdf}
%\end{minipage}
\caption{ Fermion mass dependence of the effective pion mass in QED$_2$ for a fixed $\beta=0.3$.\label{QED_mass_dep}}
\end{figure}
%
%
%In this section, we extend the study of spectroscopy with BC fermion formulation to gauge theory.  For this purpose, we  implement the BC fermion in  a 2D $U(1)$ gauge theory  and  extract the meson masses.  QED in 2D is also a confined theory and serves as a good toy model for QCD. The Gross-Neveu model, having a discrete chiral symmetry undergoes spontaneous breaking, but  in the massless Schwinger model, the chiral symmetry is continuous and cannot be spontaneously broken.
%Here we concentrate only for the lowest lying meson mass. 
Here we consider only the lowest meson state. The correlator with operator $O_1(t)$ couples to the ground state and provides the mass for the lowest  state. 
 In Fig.\ref{QED_meson}(a) we have shown the correlator at different time slices and the effective meson mass in 2D QED.  %The results are presented for the fermion (electron) mass $m_0=0.05$ and  $\beta=0.7$. 
The  Schwinger model in continuum  can be written as a  bosonic theory.   The pion mass in the bosonized theory can be exactly calculated\cite{smilga,Dur} and   for $m_0\ll g$, can be written as
 \be
 M_{eff}&=&A~ m_0^{N_f/(N_f+1)} g^{1/(N_f+1)} = A~ m_0^{5/4} g^{1/5}~~ \rm{(for~ N_f=4)},\\
 &=& A~ m_0^{4/5} \beta^{-1/10}\nonumber
 \ee
 where $A$ is a constant and $\beta=1/g^2$.   In Fig.\ref{QED_mass_dep}, we show the fermion mass dependence of the effective pion mass ($M_{eff}^{5/4}\propto m_0$). The plot is done for  small $\beta$ so that $m_0$ is always less than $ g$.  The lattice data are in
  well agreement with the analytic result.  
%   The  ground state mass $m_{eff}\approx 1.0$ and is much larger than $(2m_0)$.
%  The square of meson mass ($m_{eff}^2$) shows a linear dependence on the fermion mass as illustrated in Fig.\ref{QED_mass_dep}.
% For light fermions, the meson mass is much larger than twice the bare fermion mass $2m_0$. As the mass increases, the available phase space decreases and the contribution to the meson mass from interaction diminishes so the difference $(m_{eff}-2m_0)$ becomes smaller.
%   As can be seen from Fig.\ref{QED_mass_dep}, for heavy fermions, the meson mass becomes less than $2  m_0$.   
%    This can be explained from the fact that for heavy fermions, the quantum corrections to the effective mass become small as explained above, but the binding energy due to strong coupling is still large, so in combination of these two, the effective mass become less than the sum of individual particles as one observes for atomic or nuclear mass where the mass of the atom/nucleus is less than the sum of  the individual constituent masses.
%%%%%%%%%%%%%%
\section{Summary}
%%%%%%%%%%%%
Minimally doubled fermions may provide an  efficient lattice formalism to study chiral fermion which  is expected to be computationally cheaper than the other existing lattice formalisms. Since, both the minimally doubled fermion formulations (KW and BC) break  hypercubic symmetries on the lattice, they require non-covariant counter terms. %Only detailed numerical studies can confirm how bad or manageable  its effects are on the lattice,  and whether any meaningful computation is possible with minimally doubled fermion or not.  
In this work, we have studied  the BC fermion  in some simple models.
 We have extracted the excited state  mass  spectrum in  Gross-Neveu model using BC fermion. We have also evaluated the lowest lying  meson mass in QED$_2$.  For light fermion, the meson   is much heavier that $2 m_0$ and for heavy fermion the meson mass becomes less than $2m_0$ as the renormalized fermion mass becomes much smaller than the bare mass at strong coupling. 
  Our investigations suggest that  BC fermion formalism might be a promising  alternative to study the chiral fermions on a lattice. \\ %One obviously needs more detailed numerical study in 4D gauge theory with  dynamical BC fermion to confirm that claim.
J. G. likes to thank Stephan D{\"u}rr for his useful comments.


\begin{thebibliography}{99}
\bibitem{K} L.H. Karsten, Phys. Lett. B  {\bf 104},315 (1981).
\bibitem{W} F. Wilczek, Phys. Rev. Lett. {\bf 59},2397 (1987).
\bibitem{creutz} M. Creutz, JHEP {\bf 0804}, 017(2008);  Pos LAT2008, 080 (2008).
\bibitem{borici} A. Borici, Phys. rev. D. {\bf 78}, 074504 (2008).

\bibitem{bedaque} P. F. Bedaque, M. I. Buchoff, B.C. Tiburzi, A. Walker-Loud, Phys. Lett. B. {\bf 662},449 (2008).
%\bibitem{creutz2} M. Creutz, Pos LAT2008, 080 (2008).

\bibitem{capitani1} S. Capitani, J. Weber, H. Wittig, Phys. Lett. B681, 105 (2009).
\bibitem{capitani2} S. Capitani, M. Creutz, J. Weber, H. Wittig, JHEP {\bf 1009}, 027 (2010).
\bibitem{dc} D. Chakrabarti, S. J. Hands, A. Rago, JHEP {\bf 0906}, 060 (2009).


\bibitem{GCB}  J.~Goswami, D.~Chakrabarti and S.~Basak,
  %``Gross-Neveu model with Borici-Creutz fermion,''
  Phys.\ Rev.\ D {\bf 91}, no. 1, 014507 (2015).
 % [arXiv:1409.7999 [hep-lat]].

\bibitem{CKM} M. Creutz, T. Kimura, T. Misumi, Phys. Rev. D.{\bf 83}, 094506 (2011).
\bibitem{misumi} T. Misumi, JHEP  {\bf 1208}, 068 (2012).

%\bibitem{korzec} T. Korzec, F. Knechtli, U. Wolff, B. Leder, PoS LAT2005, 267 (2006).
%\bibitem{hands} S. J. Hands, C. Strouthos, Phys. Rev. B{\bf 78}, 165423 (2008), W. Armour, S. Hands, C. Strouthos, Phys. Rev. B{\bf 81},  125105(2010).
%\bibitem{araki} Y. Araki,  PoS LAT2011, 054 (2011); Phys. Rev B {\bf 85}, 125436 (2012).

%\bibitem{basak} S. Basak and A. K. De, Phys. lett. B {\bf 430}, 320 (1998).
%\bibitem{kimura} T. Kimura, S. Komatsu, T. Misumi, T. Noumi, S. Torii, S. Aoki, JHEP {\bf 1201}, 048 (2012).
%\bibitem{tHooft} G. 'tHooft, Nucl. Phys.  B {\bf 72}, 461, 1974.
\bibitem{Danzer}  J.~Danzer and C.~Gattringer,
  %``Excited State Spectroscopy in the Lattice Gross-Neveu Model,''
  PoS LAT {\bf 2007}, 092 (2007).
%  [arXiv:0710.1711 [hep-lat]].
\bibitem{Gutsfeld} 
  C.~Gutsfeld, H.~A.~Kastrup and K.~Stergios,
  %``Mass spectrum and elastic scattering in the massive SU(2)(f) Schwinger model on the lattice,''
  Nucl.\ Phys.\ B {\bf 560}, 431 (1999).
%  [hep-lat/9904015].
  
\bibitem{Gattringer} C. ~Gattringer et. al Phys. Lett. B {\bf 466},287 (1999).

\bibitem{cichy}  K.~Cichy, A.~Kujawa-Cichy and M.~Szyniszewski,
  %``Lattice Hamiltonian approach to the massless Schwinger model: Precise extraction of the mass gap,''
  Comput.\ Phys.\ Commun.\  {\bf 184}, 1666 (2013).
%  [arXiv:1211.6393 [hep-lat]].

\bibitem{Giusti} 
  L.~Giusti, C.~Hoelbling and C.~Rebbi,
  %``Schwinger model with the overlap Dirac operator: Exact results versus a physics motivated approximation,''
  Phys.\ Rev.\ D {\bf 64}, 054501 (2001).
%  [hep-lat/0101015].
\bibitem{Hip} W.~Bietenholz, I.~Hip, S.~Shcheredin and J.~Volkholz,
  %``A Numerical Study of the 2-Flavour Schwinger Model with Dynamical Overlap Hypercube Fermions,''
  Eur.\ Phys.\ J.\ C {\bf 72}, 1938 (2012).
%  [arXiv:1109.2649 [hep-lat]].
%\bibitem{Jiang} J. Jiang, X. Luo, Z. Mei, H. Jirari, H. Kroger, C. Wu, Phys. Rev. D {\bf 60}, 014510 (1999)

%\bibitem{hands2} S. J. Hands, A. Koci\'c, J. B. Kogut, Nucl. Phys. B{\bf 390}, 355 (1993), Ann. Phys. {\bf 224}, 29 (1993).

\bibitem{Dsource}   C.~Gattringer et. all,
  %``Derivative sources in lattice spectroscopy of excited mesons,''
  Phys.\ Rev.\ D {\bf 78}, 034501 (2008).
%  [arXiv:0802.2020 [hep-lat]].

%%%%Variational method %%%%%%%%%
\bibitem{Michael} 
  C.~Michael,
  %``Adjoint Sources in Lattice Gauge Theory,''
  Nucl.\ Phys.\ B {\bf 259}, 58 (1985).
\bibitem{Luscher} 
  M.~Luscher and U.~Wolff,
  %``How to Calculate the Elastic Scattering Matrix in Two-dimensional Quantum Field Theories by Numerical Simulation,''
  Nucl.\ Phys.\ B {\bf 339}, 222 (1990).
\bibitem{smilga} A.~V.~Smilga,
  %``Critical amplitudes in two-dimensional theories,''
  Phys.\ Rev.\ D {\bf 55}, 443 (1997).
\bibitem{Dur} Stephan Durr, Phys. \ Rev. \ D {\bf 85} 114503 (2012).
%\bibitem{Smilga}  A.V. Smilga,Phys. \ Rev.\ D {\bf 55} 443 (1997)
\end{thebibliography}
\end{document}